# Composition induced diffused to relaxor ferroelectric phase transition in lead-free (1-$x$)(Li$_{0.12}$Na$_{0.88}$)NbO$_3$-$x$BaTiO$_3$ (0 ≤ $x$ ≤ 0.40) ferroelectric ceramics

**Supratim Mitra**[*], **Ajit R. Kulkarni**

*Department of Metallurgical Engineering and Materials Science, Indian Institute of Technology Bombay, Mumbai 400076, India*

[*]*Now at: Department of Basic Sciences, NIIT University, Neemrana, Rajasthan 301705, India*

**Abstract**

(1-$x$)Li$_{0.12}$Na$_{0.88}$NbO$_3$-$x$BaTiO$_3$ (0 ≤ $x$ 0.40) ferroelectric ceramics were prepared using conventional ceramics route and their phase transitional behavior is investigated by using dielectric spectroscopy. The temperature-dependent dielectric permittivity $\varepsilon'(T)$ shows a diffused ferroelectric-paraelectric transition for all compositions. An acceptable and competent characterizing parameter ($D$) of diffused phase transition (DPT), defined by Uchino *et al.* [J Am Ceram Soc 2010;93:4011], was measured and validated. Interestingly, a crossover from diffused ferroelectric phase transition (FE-DPT) to relaxor ferroelectric (RFE) transition is found for the composition $x$ ≥ 0.225. The FE-DPT is characterized by a frequency-independent temperature of dielectric maxima ($T_m$), while a RFE is found to have frequency-dependent $T_m$ satisfying Vogel-Fulcher relation. The composition induced crossover is attributed to the dynamics of different PNR size and relaxation times that varies with different BaTiO$_3$ content ($x$) leading to the appearance of a FE-DPT or RFE.

# 1. Introduction

Ferroelectrics with diffused phase transition (DPT) including relaxors have received much attention in recent years both for fundamental research and technological applications **[1-3]**. Relaxor ferroelectrics (RFEs) have been extensively studied for their exceptionally higher dielectric constant and attractive piezoelectric properties, which are essential for high energy density capacitors and actuators. Therefore, from the technological aspect it is very important to have a simple empirical formula which would be able to explain the temperature dependence of dielectric permittivity near transition and quantitatively describes the degree of diffuseness of a diffuse phase transition (DPT) in ferroelectric materials including RFEs. Lead-based RFEs have been widely investigated for last couple of decades with substantial interest for their particular characteristics **[4]**, however the attention has been moved towards lead-free ferroelectric materials due to environmental concern of toxic lead (Pb) that contained in lead-based materials.

In RFE systems, the dynamic nature of nano-sized cation ordered polar nano regions (PNRs) dispersed in a disordered matrix, results in relaxor behavior. A typical temperature-dependent dielectric permittivity shows broad peak that shifts towards higher temperature with frequencies, which could be further substantiated by close agreements with Vogel-Fulcher (VF) relation. A frequency dispersion on the low-temperature slope of the dielectric peak has also been observed as characteristics of relaxor behavior. Many theoretical models have been developed to explain origin of relaxor behavior which includes super-paraelectric model **[3]**, glass-like freezing of PNRs **[5]**, random field model **[6]** etc. On the contrary, there are many ferroelectric systems such as, $(Pb,La)TiO_3$ **[7]**, $(Pb,La)(Ti,Zr)O_3$, $Pb(Sc,Ta)O_3$, $(Pb,Ca)TiO_3$, $(Sr,Ba)NbO_3$, $(K,Sr)TiO_3$, $(Ba,Ca)TiO_3$, $(Ba,Sn)TiO_3$, $Ba(Ti,Sn)O_3$ **[8-10]**, $(Ba,La)(Ti,Al)TiO_3$ **[11]**, $(Ba,La)(Ti,Ga)TiO_3$ **[12]**, $Ba(Zr,Ti)O_3$ **[13]**, $(Ba,Ca)(Ti,Zr)O_3$ **[14]**, $(Li,Na)NbO_3$ **[15]** which show a frequency dispersion, broad dielectric peak at the frequency-independent peak temperature $T_m$ and therefore could not be recognized as typical RFEs. These materials could be classified as ferroelectrics with diffused ferroelectric phase transition (FE-DPT). The broadening of the transition peak represents degree of diffuseness, an important characteristic of diffused ferroelectric phase transition (DFPT). The true origin of DFPT is ambiguous. However it is believed to be mainly associated with defects in the materials, arising due to compositional fluctuation, grain configuration, cation disorder, point defects and microscopic heterogeneities [16]. In ferroelectrics, especially with a diffused ferro-paraelectric phase transition, the relaxation processes of the defect species might overlap with the intrinsic polarization contribution to the total dielectric permittivity and thus shape the overall dielectric response and make the phase transition more diffuse [17, 18].

Lithium sodium niobate, (Li,Na)NbO$_3$ (LNN) ceramics, an important materials system for filter and resonator applications, showed a diffused ferroelectric phase transition (DFPT) [15]. The DFPT in LNN system arises mainly due to compositional disorder, i.e., disorder induced due to occupancy of an equivalent site in the lattice by different types of cations. It is also possible that different concentration of substituting ions leads to different degrees of compositional disorder. Consequently, these solid solutions become interesting because of nature of phase transition under the influence of substitutional effect [19]. The solid solution of LNN with BaTiO$_3$ (BT) i.e., $(1-x)$Li$_{0.12}$Na$_{0.88}$NbO$_3$-$x$BaTiO$_3$ ($0 \leq x$ 0.40) showed excellent electrical properties in the vicinity of morphotropic phase boundary (MPB) composition and an interesting evolution in phase transitional behavior with increasing BaTiO$_3$ content [20]. In this report, $(1-x)$LNN-$x$BT solid-solutions were prepared using conventional ceramic processing route and their nature of phase transitional behavior is investigated in detail by using dielectric spectroscopy. The observed diffuse phase transition in these systems has been studied quantitatively based on recently developed phenomenological model by Uchino *et. al.*[4]. The influence of structure, grain configuration and defects on the origin of diffused phase transition in (1-$x$)LNN-$x$BT ceramics are also discussed. Composition induced diffused to relaxor ferroelectric phase transition is observed as BT content ($x$ mol%) is increased. The RFE behavior, characterized by Vogel-Fulcher relations and relaxation dynamics of PNRs for an optimal BT content ($x \geq 0.225$) which is attributed to a critical size of PNRs to bring in the relaxor behavior, have been studied.

## 2. Phenomenological theory of diffuse ferroelectric phase transition (DFPT)

Smolensky [21] brought in first the famous microscopic compositional fluctuation model to characterize the diffuseness parameter. The model considered that there exist microscopic regions of slightly different compositions with different Curie temperatures ($T_C$). The temperature characteristic of dielectric permittivity is resulting from the distribution of Curie-temperatures ($T_C$) of these individual micro-volumes and can be expressed as Gaussian type:

$$f(T_C) = \exp[-(T_C - T_m)^2/2\delta^2] \tag{1}$$

where, $T_m$ is the average Curie-temperature of the ferroelectric material, $\delta$ is the standard deviation of the Gaussian function and defined as the degree of diffuseness. A power series expansion of the above equation and neglecting the higher order terms, an approximate quadratic relation between dielectric constant $\varepsilon$ and temperature T can be readily obtained as,

$$1/\varepsilon = [1+(T_C - T_m)^2/2\delta^2]/\varepsilon_m \tag{2}$$

where, $\varepsilon_m$ is the value of maximum permittivity at the phase transition temperature $T_m$ and $\delta$ is used as the measure of diffuseness which must be adequate to validate the above equation. Since, $\delta$ is found to be temperature and frequency dependent in many studies [22-26], the deviation of experimental data from the above quadratic equation (Eqn. 2) motivated many authors to search for new formulation. Uchino and Nomura **[23]** modified the equation and suggested an empirical relation between ε and T as,

$$1/\varepsilon = [1+(T_C - T_m)^\gamma/2\delta^2]/\varepsilon_m \ , \ 1 \leq \gamma \leq 2, T > T_m \tag{3}$$

where, γ is used instead of 2 and defined as diffuseness parameter which has a value 1 for ideal ferroelectrics and 2 for a complete diffused phase transition (DPT) or classical RFE. The value of γ is obtained from the slope of the linearly fitted curve of $\ln(1/\varepsilon - 1/\varepsilon_m)$ vs $\ln(T-T_m)$ **[26, 27]** and therefore, found to be different in different temperature regions in ε-T curve. These different values of γ when calculated for the different temperature regions for the same system e.g., $(Pb,Mg)NbO_3$ ceramics **[24, 28]** leads to ambiguities in the model. Furthermore, γ-value has also been found to deviate from the range $1 \leq \gamma \leq 2$ for $(Pb,Mg)NbO_3$, $BaTiO_3$, $Ba(Zr,Ti)O_3$ **[29-31]**, $(Pb,Ca)TiO_3$, $Pb(Fe,W)O_3$-$PbTiO_3$ **[10, 32]** ceramics systems. Moreover, this model could not be used when $T < T_m$, while DPT describes the characteristics both below and above $T_m$. Therefore, it can be concluded that this model is not suitable for materials with a broad dielectric peak in temperature-dependent permittivity curve. Then, Lobo *et al.* **[16]** suggested a complex equation between ε and T over the whole temperature range

$$1/\varepsilon(T) = \int_0^\infty \varepsilon^{-1}(T_C) \exp\left[-\frac{(T_m - T_C)^2}{v^2}\right] \frac{1}{v\sqrt{\pi}} dT_C \tag{4}$$

where, $\varepsilon^{-1}(T_C)$ is the inverse permittivity of each micro-volumes and v is defined as diffuseness degree. However, this model has also been found not suitable for the entire range of temperature studied by Lobo *et al.* for $BaTiO_3$ **[16]**. There are few other methods which were tried for characterizing the effective diffuseness such as full width at half-maxima, width at nine-tenth at maxima, etc. in ε-T curve for $(Pb,Mg)NbO_3$-$PbTiO_3$ **[26]**, $(Ba,Ca)TiO_3$ **[33]** ceramics systems, however most of them were unproven. On the contrary, despite their obvious shortcomings, equation (2) and (3) remain the most consented formulae to describe the diffuseness.

Recently, a reasonable and effective diffuseness parameter of DPT was defined by Uchino *et al.* based on Smolensky's microscopic composition fluctuation model and Devonshire's phenomenological theory [4]. In this model, the characterizing parameter of diffuseness degree, $D$ is associated with the temperature interval where volume of polar microscopic regions change due to appearance of new polar microscopic regions, and defined as,

$$D = T_{(\partial \varepsilon(T)/\partial T)_{max}} - T_{(\partial \varepsilon(T)/\partial T)_{min}} \tag{5}$$

Where, $T_{(\partial \varepsilon(T)/\partial T)_{max}}$ and $T_{(\partial \varepsilon(T)/\partial T)_{min}}$ are the temperatures where $\partial \varepsilon(T)/\partial T$ reach maxima and minima respectively. The temperature interval, $D$ reflects the degree of diffuseness macroscopically and can be calculated easily for different systems. This parameter proved to be reasonably effective and found to have good universality for various ferroelectric materials [4].

## 3. Experimental

Solid solution of $(1-x)\text{Li}_{0.12}\text{Na}_{0.88}\text{NbO}_3$-$x\text{BaTiO}_3$ [$(1-x)$LNN-BT] ($0 \leq x$ 0.40) were prepared using conventional ceramic processing route. The details of synthesis of $(1-x)$LNN-BT ceramic powders and preparation of sintered pellets were reported elsewhere [20]. The phase purity of the ceramic samples was checked by powder X-Ray Diffraction (XRD) at room temperature using X-ray diffractometer (X'Pert, PANalytical) with Cu-Kα radiation. For electrical measurements, all the pellet samples were polished and silver paste were fired on both the surfaces as electrodes. The dielectric permittivity of the unpoled samples was recorded using Impedance Analyzer (Alpha High Resolution, Novocontrol, Germany) in the frequency range 0.01 Hz-1 MHz over the temperatures -100-100 °C and 50-500 °C using low- and high-temperature sample holder respectively.

## 4. Results and analysis
### 4.1 Quantitative description of diffuse phase transition

Figure 1(a) shows the temperature-dependent permittivity, $\varepsilon'(T)$ of (1-$x$)LNN-$x$BT ceramics for the composition range $0 \leq x \leq 0.30$ at 1 kHz. All the compositions show a diffused ferroelectric-paraelectric transition peak. However, quantitatively the degree of diffuseness, $D$ has been calculated from $\varepsilon'(T)$ plot (Fig. 1(a)) using Eqn. (5) and are listed in Table 1 and plotted in figure 2 as well. The $D$-value could not be calculated for $x = 0.40$ as $T_{(\partial \varepsilon(T)/\partial T)_{min}}$ become less than -100 °C (the lowest measured temperature). It can be seen from figure 2 and the data given in Table 1 that the $D$-value first decreases from 62 to 25 as $x$ increases from 0.0 to 0.10 and then gradually increases to 100 as $x$ increases up to $x = 0.30$. The range of $D$-values, that obtained in (1-$x$)LNN-$x$BT ceramics are comparable with systems discussed in the report by Uchino *et. al.* **[4]** where the effectiveness of the newly introduced parameter, $D$ was validated.

As mentioned in section 2, the degree of diffuseness is generally measured from the diffuseness parameter $\gamma$ using Eqn. (3), however it may sometimes cause contradiction in its measured values ($1 \leq \gamma \leq 2$). Hence, it is obvious to check whether the contradiction exist in (1-$x$)LNN-$x$BT ceramics. The $\gamma$-values are also calculated and the obtained values are listed in Table 1. There is no significant difference found in the trend of diffusivity ($\gamma$) and diffusivity degree ($D$) that changes with BT content ($x$ mol%) as can be observed in figure 2. However, the $\gamma$-value for the composition $x = 0.40$ is found to be 2.05 indicating the possibilities of presence of some ambiguity. Therefore, it confirms that parameter $D$ is more effective and reasonable in characterizing the diffuseness degree without any ambiguity. The $\gamma$-values for the compositions that are close to 2 are significance of an almost complete diffused phase transition characteristics of ferroelectric with diffuse phase transition (FE-DPT) or RFEs. Therefore, they do not follow Curie-Weiss law and a deviation can be observed when inverse of dielectric permittivity ($1/\varepsilon'$) is plotted against temperature ($T$) as seen in figure 1(b). The deviation from Curie-Weiss law, which is also a measure of degree of diffuseness is defined as, $\Delta T = T_B - T_m$ (where, $T_B$ is the Burns-temperature above which materials follow Curie-Weiss law, often considered as the onset temperature of local polarization [34, 35] and $T_m$ is the temperature of maximum dielectric permittivity) and the obtained $\Delta T$, $T_m$ and $T_B$ –values are listed in Table 1. The variation of $\Delta T$ with $x$ is also shown in figure 2 and found consistent with the other diffuseness parameters ($\gamma$, $D$).

**4.2 Diffused ferroelectric phase transition in (1-$x$)LNN-$x$BT Ceramics (0 ≤ $x$ ≤ 0.20)**

Figure 3 depicts the variation of real ($\varepsilon'$) and imaginary ($\varepsilon''$) parts of dielectric permittivity of (1-$x$)LNN-$x$BT ceramics for $0 \leq x \leq 0.20$ as a function of temperature ($50 \leq T \leq 500$ °C) at different frequencies (0.5-260 kHz). Both $\varepsilon'$ and $\varepsilon''$ exhibit diffused ferroelectric to paraelectric phase transition peak near $T_m$. In addition to the peak at $T_m$, another weak anomaly appears as a shoulder at the lower temperature side of permittivity maxima as appeared in pure LNN ceramics [15]. In (1-$x$)LNN-$x$BT ceramics, this anomaly is seen to shift towards lower temperature side and weaken as BT content ($x$) increases and finally disappears for $x > 0.15$. The observed higher values of $\varepsilon'$ at very low frequencies are mainly attributed due to the extrinsic space charge polarization arising from the grain boundaries and sample-electrode interfaces. A sharp increase in $\varepsilon''$ above ferro-paraelectric transition ($T_m$) is observed which is strongly frequency-dependent and attributed due to an increase in conductivity arising from an interaction between free carriers (electron and holes) with grain boundary potential barriers. The temperature corresponds to the maximum in $\varepsilon'(T_m)$ is found almost at the same temperature to that of $\varepsilon''(T_m)$ for compositions $0 \leq x \leq 0.20$ in (1-$x$)LNN-

$x$BT ceramics. However, a shift in $T_m$ towards higher temperature side with increasing measuring frequencies in $\varepsilon'(T)$ plots has not been observed. All such behavior is associated with the diffuse phase transition (DPT) in ferroelectric systems and hence one can simply rule out the relaxor ferroelectric transition being responsible for DPT. The diffuse character in the ferroelectric transition is believed to arise due compositional fluctuation **[21, 36, 37]** as described in section 4.2. The presence of microregions with local composition varying from the average composition over a small length scale and their transformation with a Gaussian distribution of Curie temperature results in a diffused phase transition.

**4.3 Relaxor Behavior in (1-$x$)LNN-$x$BT Ceramics (0.225 ≤ $x$ ≤ 0.40)**

Figure 4 depicts the variation of real ($\varepsilon'$) and imaginary ($\varepsilon''$) parts of dielectric permittivity of (1-$x$)LNN-$x$BT ceramics for $x$ = 0.225 (50 ≤ $T$ ≤ 500 $^o$C) and 0.25 ≤ $x$ ≤ 0.40 (-100 ≤ $T$ ≤ 100 $^o$C) as a function of two different temperature ranges respectively and at different frequencies (0.5-260 kHz). In addition to a DPT, RFEs are characterized by shifting of temperature of dielectric maxima ($T_m$) towards higher temperature with increasing measuring frequencies and frequency dispersion in the temperature-dependent dielectric permittivity as can be seen in figure 4 for (1-$x$)LNN-$x$BT ceramics (0.225 ≤ $x$ ≤ 0.40). The temperature corresponds to the maximum in $\varepsilon'(T_m)$ is found at the higher temperature to that of $\varepsilon''(T_m)$ which is also another criteria for being a RFE.

Unlike normal ferroelectrics, RFE have non-zero polarization at temperature well above $T_m$ due to the presence of micro or nano polar regions (PNRs) and is widely believed to be responsible for relaxor behavior **[3, 5, 36, 38-41]**. As stated earlier that relaxors are compositionally disordered on an atomic scale and therefore form quenched local dipoles and local random field which exists even in the cubic paraelectric (PE) phase. Upon cooling the PE phase transforms into an ergodic relaxor (ER) phase as a result of local phase transition and the PNRs start appearing at temperature $T_B$ (usually $\gg T_m$), the so called Burns temperature (given in table 1). The dynamic PNRs grow in size and start interacting with each other and the interaction become stronger near $T_m$ due to instability of phases. On further cooling, the interaction between these PNRs slows down their dynamics and freezes near $T_f$ where the size of the PNRs has increased significantly and gradually transforms into a non-ergodic relaxor (NR) phase.

### 4.3.1 Vogel-Fulcher (VF) relation

An empirical Vogel-Fulcher (VF) relation can be used to account the frequency dependence of temperature maxima of dielectric relaxation peak ($T_m$) in RFEs. According to which, the polarization switching frequency $f$ is related to the activation energy for polarization fluctuations of an isolated cluster as,

$$f = f_o exp[-E_a/k_B(T_m-T_f)] \tag{6}$$

where, $T_f$ is the freezing temperature and $f_o$ is the pre-exponential factor. Figure 5 shows the plot between $\ln(f)$ and $T_m$ for (1-$x$)LNN-$x$BT ceramics in the composition range $x$ = 0.225-0.40. The dotted points represent experimental data and the solid line is the non-linear VF fit to eqn. (6). The plots reveal an excellent fit to VF relation characterizing a relaxor behavior. However, it is worth mentioning that a fitting of VF relation for the composition, $x < 0.225$ was not successful and thereby confirming a DFPT. The summary of phenomenological parameters of VF relation for $0.225 \leq x \leq 0.40$ is listed in table 2. The relaxation behavior of PNRs in (1-$x$)LNN-$x$BT ceramics in the composition range $x$ = 0.225-0.40, is investigated with the help of $\varepsilon''(f)$ and $M''(f)$ plot at selected temperature regions for $x$ = 0.30 as an representative composition.

**4.3.2 Dielectric relaxation behavior in (1-$x$)LNN-$x$BT Ceramics ($x$ = 0.30)**

In general, the origin of appearance of PNRs below $T_B$ upon cooling are not well studied because of measurement limitations (as $T_B$ is formed at very high temperature for Pb-based RFEs and frequency range of dielectric dispersion falls out of experimental regime) **[42]**. In (1-$x$)LNN-$x$BT ceramics, the composition range $0.225 \leq x \leq 0.40$ shows RFEs whose $T_B$ ranges from 140 to 35 $^{o}$C respectively. The PNRs are believed to be in dynamic state for $T_f \leq T \leq T_B$ and a relaxation processes takes place. Figure 6 shows $\varepsilon''(f)$ and $M''(f)$ plot at selected temperature regions (-50-200 $^{o}$C) for $x$ = 0.30 as an representative composition. A major peak in $M''(f)$ plot at low frequency is observed as bulk response followed by a shoulder in the frequency range of 10-10$^5$ Hz as seen in figure 6(a) and 6(b). The bulk capacitance ($C_b$) was calculated from $M''_{max} = \varepsilon_0/2C_b$ and found close to sample capacitance, however the capacitance for the shoulder is found an order higher than that of the bulk which is attributed to the presence of a highly polarizable small dimension entity **[43]**. This is an evidence of relaxation of PNRs which present in the materials and is in dynamic in the temperature range of -27 = $T_f \leq T \leq 100 = T_B$. The presence of the shoulder in this particular temperature regime and the disappearance beyond the mentioned temperature range as revealed in figure 6(b). The same PNRs relaxation characteristics are also observed in $\varepsilon''(f)$ plot where the shoulder appears in the temperature range of -27 to 100 $^{o}$C and the same frequency range of 10 Hz-10$^5$ Hz as seen in figure (a) and (b). This is evident that PNRs are formed near $T_B$ (as the appearance of shoulder) and shows a relaxation processes and their dynamics freezes below $T_f$ (as disappearance of shoulder). The temperature range ($T_f \leq T \leq T_B$) found from the relaxation of PNRs is consistent with the results for $T_f$ and $T_B$ found using VF relation (sec. 4.4.1) and deviation from Curie-Weiss law (sec. 4.1) respectively. The similar results are also obtained for other compositions.

## 5. Discussion

Let us now discuss the ferroelectric-paraelectric phase transition behavior based on temperature-dependent permittivity $\varepsilon'(T)$ data and composition-induced crossover from FE-DPT to RFE took place in (1-$x$)LNN-$x$BT ceramics for the studied composition range $0.0 \leq x \leq 0.40$.

### 5.1 Diffused phase transition behavior

Interestingly, all the compositions show a diffuse phase transition (DPT) and therefore, it is always desirable to measure the degree of diffuseness quantitatively using phenomenological model for technological aspects. Here, in this study we have used recently developed phenomenological model by Uchino *et. al.* **[4]** to measure the degree of diffuseness ($D$) and has been compared with the other commonly used model (eqn. (2) and (3)), in which it is found $D$ is more reasonable and effective parameter to measure for (1-$x$)LNN-$x$BT ceramics. The origin of this DPT behavior in (1-$x$)LNN-$x$BT ceramics have been tried to explain in the light of structural distortion, inhomogeneous grain size distribution and defects that present in the materials.

### 5.1.1 Effect of structural distortion

The origin of diffuseness in dielectric-permittivity peaks is mainly due to compositional fluctuation and structural disorder in the arrangements of cations in one or more crystallographic sites. In (1-$x$)LNN-$x$BT ceramics, $A$-site is occupied by $Ba^{2+}$ (ionic radius (IR) = 1.47 Å, for coordination number (CN) = 9) and $Na^+$ (IR = 1.24 Å, for CN = 9)/$Li^+$ (IR ~ 1.00 Å, for CN = 9) whereas $B$-site is occupied by $Ti^{4+}$ (IR = 0.605 Å, for CN = 6) and $Nb^{5+}$ (IR = 0.64 Å, for CN = 6) in the perovskite $ABO_3$ structure. Therefore, DFPT behavior should be attributed to the cationic disorder induced by both $A$ and $B$-site substitutions. The measuring parameters of diffusivity, γ and *D* are found to have a decreasing tendency as BT content (*x*) approaches to *x* = 0.10 (MPB composition) **[20]** as shown in Figure 2. This could be due to cationic ordering effect which minimizes the composition fluctuation. As the two structural phases, namely orthorhombic and tetragonal phases, coexists at this composition range the compositional fluctuation is minimum as the degree of random occupancies of different cations decreases resulting in a sharp phase transition **[44]** however, need not be always true **[10]**. Also, if the difference between ionic radii of different ions in each cation site is large, it favors the *A* or *B*-site ordering degree and local compositional fluctuation may decrease which leads to a sharp phase transition **[45-48]**. In case of (1-*x*)LNN-*x*BT ceramics the difference in ionic radii at *B*-site is very small whereas, there is a large difference in ionic radii at *A*-site, and could be responsible for decrease in diffusivity (γ, *D*) in dielectric peaks near *x* = 0.10. For BT content, *x* > 0.10, the DFPT could be attributed to the structural disorder in tetragonal-rich phase due to multiple cation occupancies.

**5.1.2 Effect of grain size**

DPT may also arise due to internal stress in the grain **[30]**. In smaller grains, because of the higher concentration of grain boundaries, internal stress occurs to a greater extent. Furthermore, when a material changes from paraelectric to ferroelectric phase upon cooling from higher to lower temperature, the internal stress is compensated by the volume fraction of long-ranged ferroelectric domains. If the grain sizes are small (as compared to ferroelectric domain size) this compensation is not fully completed leading to different Curie-temperature for different grains which results in a diffused phase transition. Therefore, grain size/distributions is found reasonably an important influencing factor for the change of a delicate balance between the long range and short range forces **[13, 30]** and on the diffuse character of dielectric permittivity peak. In our recent report for (1-*x*)LNN-*x*BT ceramics

**[20]**, showed bimodal grain-size distribution, consisting of a large fraction of smaller grains and a small fraction of larger grains. This character of bimodal distribution is found to increase as $x$ (BT content) increases from 0.05 to 0.125 and then gradually become uniform as $x$ increases from 0.125 to 0.40. Consequently, the average grain size, $G$ has also increased from 2.1 to 6.5 μm as $x$ increased from $x = 0.05$ to 0.125 and then gradually decreased to 2.1 μm as $x$ reaches to 0.40, as seen in Table 1. Therefore, in coarse-grained $(1-x)$LNN-$x$BT ceramics the values of measuring parameters of diffusivity, $\gamma$ and $D$ have lower values (Fig. 2) compared to fine grained which is in agreement with other lead-free titanate-based piezoelectric system **[16, 49-55]**.

**5.1.2 Effect of defects**

The dielectric response in these materials is arising from various contributions which include conduction and dipolar relaxation processes that are often attributed to defects. Each of these contributions becomes significant at particular frequency/temperature interval. However, not all these microscopic mechanisms are well understood, the total dielectric permittivity are believed to be mainly arising from intrinsic polarization and carrier induced polarization **[17]**. These contributions increase near transition temperature leading to the fact that at transition region one of the transverse optical phonon modes weakens due to certain coupling between them. The carrier polarization dominate at low frequency and become less dominant at higher frequencies giving rise a broad transition peak. The combined effects modify the total contribution to the dielectric permittivity results in a broad peak. In $(1-x)$LNN-$x$BT ceramics, the defect reaction can be written as **[43]**,

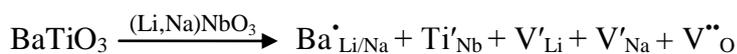

$$BaTiO_3 \xrightarrow{(Li,Na)NbO_3} Ba^{\bullet}_{Li/Na} + Ti'_{Nb} + V'_{Li} + V'_{Na} + V^{\bullet\bullet}_{O}$$

Therefore, it is believed that the cations and oxygen vacancies influence the overall dielectric response and make the phase transition more diffuse.

## 5.2 Diffused to relaxor ferroelectric phase transition

As characteristics of DPT behavior, the materials are found to start deviating from Curie-Weiss law at temperature $T_B$ upon cooling known as the onset temperature of local polarization. The quantitative measure of the deviation from Curie-Weiss law is $\Delta T = T_B - T_m$ ($T_m$ is the temperature of maximum dielectric permittivity). All these quantitative measuring parameters of diffuseness ($D$, $\gamma$, $\Delta T$) that used in this study are found to decrease with BT content ($x$) up to $x = 0.10$ and then increase almost linearly with further increase in $x$ indicating an almost completely diffused phase transition characteristics of FE-DPT or a RFE. However, all of them do not show a RFE behavior. The RFEs are mainly differing from FE-DPT in terms of frequency-dependent $T_m$ which could be characterized by matching with Vogel-Fulcher (VF) relation, which are here used as criterion of crossover from FE-DPT to RFEs.

The RFE behavior is believed to arise from the dynamics of PNRs that stars to grow in the paraelectric (PE) phase at $T_B$ ($>> T_m$) up on cooling from high-temperature due to local phase transition (condensation of phonon soft mode) and keep on growing with further decrease in temperature. The dielectric dispersion observed in the vicinity of $T_m$ (below and above $T_m$) is a result of interaction among the PNRs, however prevented by thermal disordering effect. As temperature decreases thermal disordering effects are lowered and interaction becomes stronger. On further cooling, the number of PNRs decrease abruptly at the freezing temperature $T_f$ as they merge into larger size and result in slowing down their dynamics and stop relaxing. In general, the relaxation behavior could not be possible to study as $T_B$ is formed at very high temperature for Pb-based RFEs and frequency range of dielectric dispersion falls out of experimental regime. However, in (1-$x$)LNN-$x$BT ceramics for $x = 0.30$ relaxation behavior was possible to investigate using $\varepsilon''(f)$ and $M''(f)$ plot at selected temperature regions.

The composition-induced crossover from FE-DPT to RFE is mainly arising from relaxation of PNRs characterized by VF relation. In (1-$x$)LNN-$x$BT ceramics, it is seen that at higher concentration of BT content ($x$), $T_m$ starts following VF relation and a successful fit could be obtained only for $x \geq 0.225$ indicating a RFEs and below this composition it is a FE-DPT. The nature of this different phase transitions in FE-DPT and RFE is still a matter of discussion, however, this could be explained qualitatively in terms of different PNR sizes and their relaxation time **[56]**. The compositions (0.225 $\leq x \leq$ 0.40) with RFE are more compositionally disordered and the size of newly grown PNRs is determined by the correlation length of the phase transition order parameter (i. e., polarization). The PNR sizes of the RFEs are small as their correlation length is small resulting a relaxation frequency that falls in the accessible measurement frequency range. However, the compositions ($x <$ 0.225) with FE-DPT are less compositionally disordered and the size of the newly grown PNRs are so huge that they took longer time to relax or the relaxation frequency are so small that they do not fall in the accessible measurement frequency range **[19]**.

## 6. Conclusion:

In conclusion, the dielectric spectroscopic analysis of temperature and frequency dependent dielectric permittivity data with the use of phenomenological model, gives a quantitative description of DPT in (1-$x$)LNN-$x$BT ceramics. The solid solution with an optimal BT content, $x \geq$ 0.225 exhibits RFE behavior with frequency-dependent peak temperature $T_m$, satisfying Vogel-Fulcher relations. The relaxation dynamics of PNRs which bring in the relaxor behavior in (1-$x$)LNN-$x$BT ceramics for $x \geq$ 0.225 is attributed to a critical size of PNRs.

**Figure Captions:**

**Figure 1:** Temperature dependence of (a) dielectric permittivity (ε) and (b) inverse dielectric permittivity (1/ε) measured at 1 kHz for (1-$x$)LNN-$x$BT ceramics (0 ≤ $x$ ≤ 0.40).

**Figure 2:** Variation of diffusivity (γ) and degree of diffusivity ($D$) with BT content ($x$ mol%) in (1-$x$)LNN-$x$BT ceramics (0 ≤ $x$ ≤ 0.40).

**Figure 3:** Temperature dependence of real and imaginary part of dielectric permittivity measured at various frequencies (0.5-260 kHz) of (1-$x$)LNN-$x$BT ceramics for $x$ = 0.0-0.20.

**Figure 4:** Temperature dependence of real and imaginary part of dielectric permittivity measured at various frequencies (0.5-260 kHz) of (1-$x$)LNN-$x$BT ceramics for $x$ = 0.225-0.40.

**Figure 5:** Plot of frequency vs temperature of maximum dielectric permittivity. The open spheres represent the experimental data and the solid line is the fit to Vogel-Fulcher relation.

**Figure 6:** Variation of imaginary part of modulus (M″) and dielectric permittivity (ε″) with frequency for the temperature range of (a) and (c) -50 to100 °C; (b) and (d) 100 to 200 °C for (1-$x$)LNN-$x$BT Ceramics ($x$ = 0.30).

**Table Captions:**

**Table 1:** Variation of degree of diffusivity ($D$), diffusivity (γ), temperature of maximum dielectric permittivity ($T_m$), Burns-temperature ($T_B$) and temperature deviation from ideal Curie-temperature (Δ$T$) with BT content ($x$ mol%) in (1-$x$)LNN-$x$BT ceramics (0 ≤ $x$ ≤ 0.40).

**Table 2:** Summary of parameters of Vogel-Fulcher relation for (1-$x$)LNN-$x$BT ceramics for $x$ = 0.225-0.40

**TABLE 1:**

| $x$ (mol%) | $D$ (°C) | $\gamma$ | $T_B$ (°C) | $T_m$ (°C) | $\Delta T$ (°C) | $G$ (µm) |
|---|---|---|---|---|---|---|
| 0 | 62 | 1.72 | 375 | 345 | 30 | - |
| 5 | 37 | 1.51 | 345 | 320 | 25 | 2.11 |
| 10 | 25 | 1.43 | 275 | 265 | 10 | 5.10 |
| 15 | 35 | 1.45 | 230 | 207 | 23 | 6.32 |
| 20 | 48 | 1.57 | 150 | 125 | 25 | 6.51 |
| 22.5 | 58 | 1.74 | 140 | 89 | 51 | 4.21 |
| 25 | 66 | 1.87 | 135 | 56 | 79 | 3.80 |
| 27.5 | 74 | 1.92 | 115 | 6 | 109 | 3.36 |
| 30 | 100 | 1.97 | 100 | -15 | 115 | 2.62 |
| 40 | -- | 2.05 | 35 | -90 | 125 | 2.10 |

**TABLE 2**

| $x$ (mol %) | $T_f$ (°C) | $E_a$ (eV) | $f_0$ (Hz) | $R^2$ |
|---|---|---|---|---|
| 0.225 | 77 | 0.0344±0.0101 | 2.04x10$^{12}$ | 0.976 |
| 0.25 | 48 | 0.0155±0.0052 | 4.05x10$^{8}$ | 0.997 |
| 0.275 | -2 | 0.0287±0.0153 | 4.40x10$^{10}$ | 0.999 |
| 0.30 | -27 | 0.0289±0.0280 | 7.90x10$^{10}$ | 0.978 |
| 0.40 | -122 | 0.0325±0.0280 | 3.80x10$^{11}$ | 0.998 |

**FIGURE 1:**

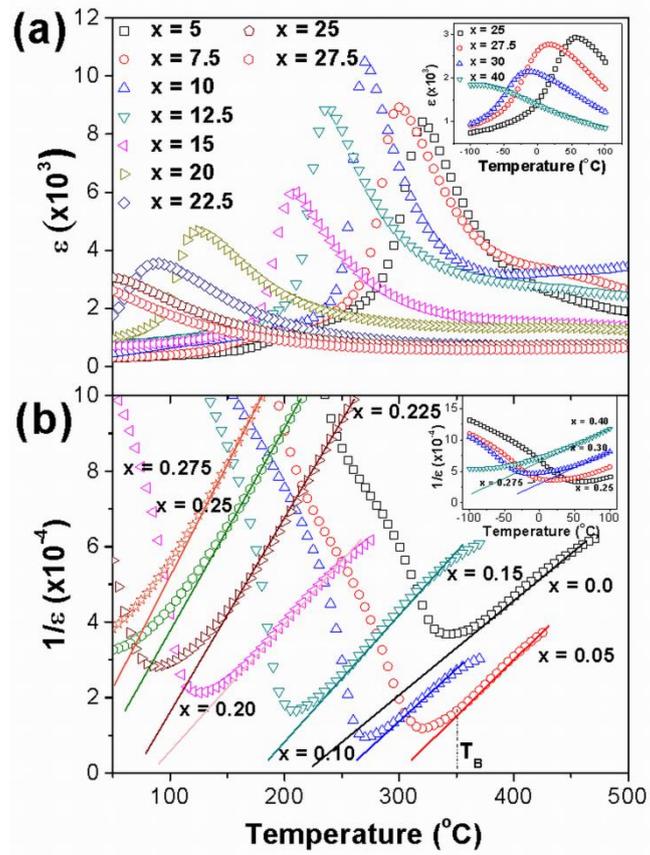

**FIGURE 2:**

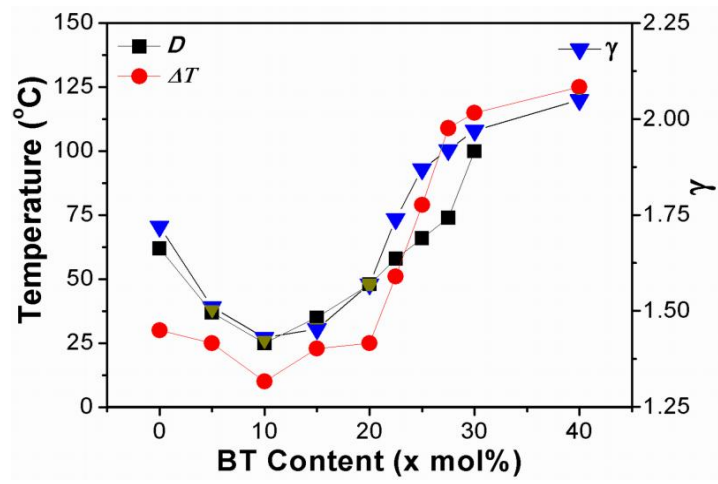

**FIGURE 3:**

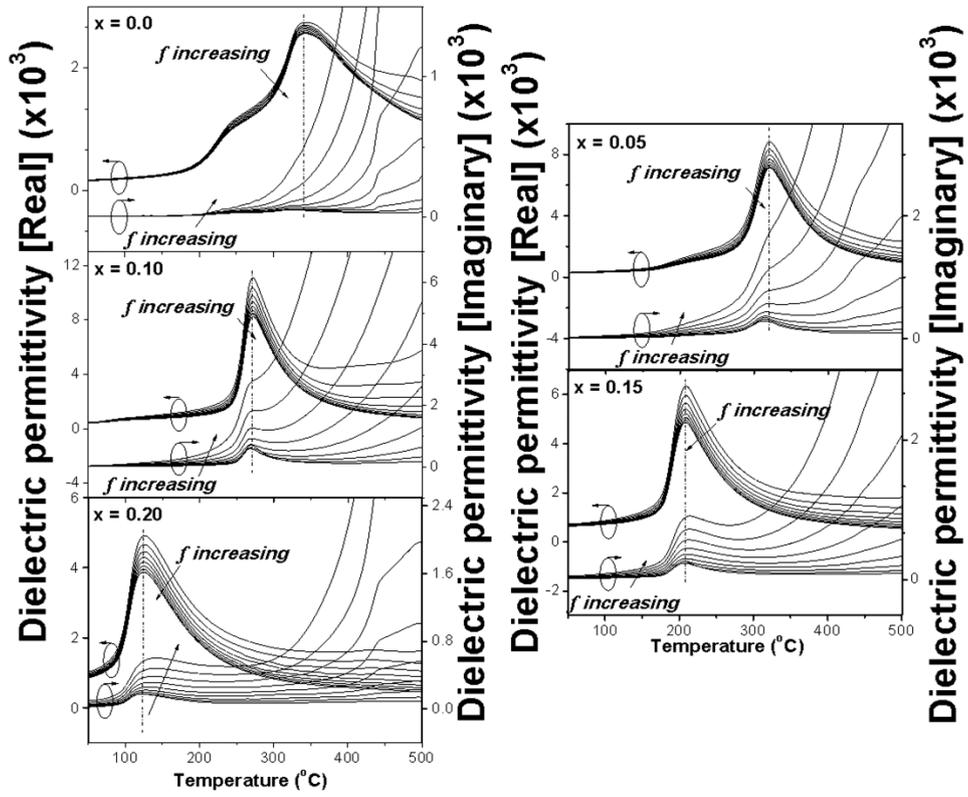

**FIGURE 4:**

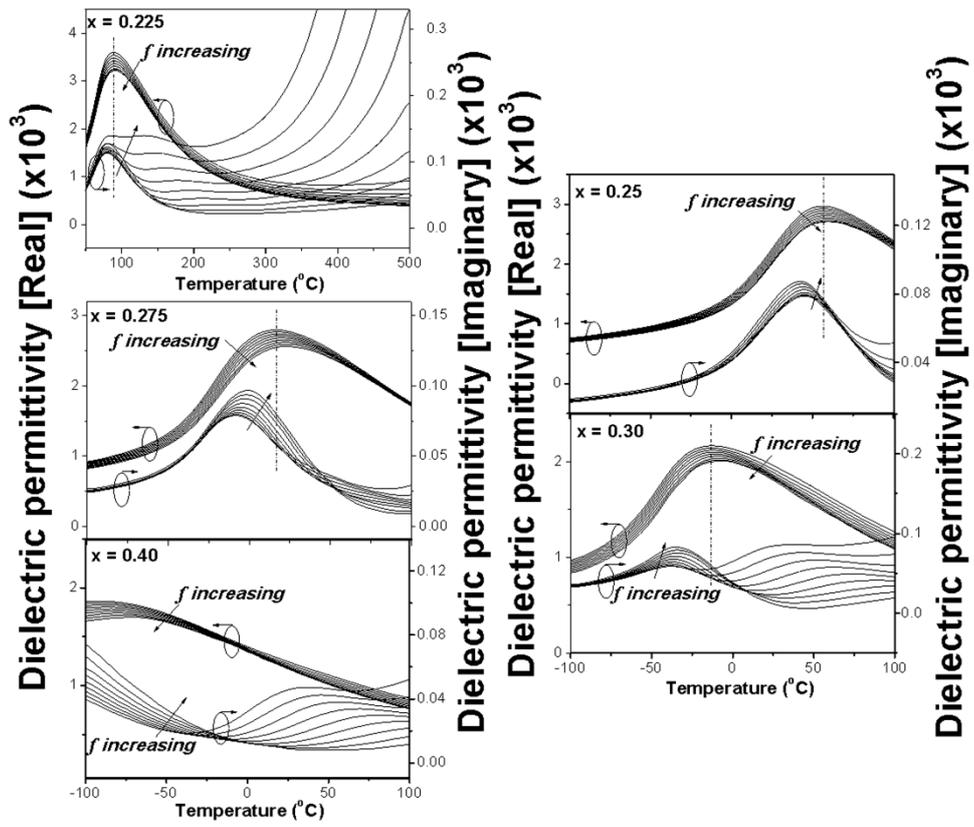

**FIGURE 5:**

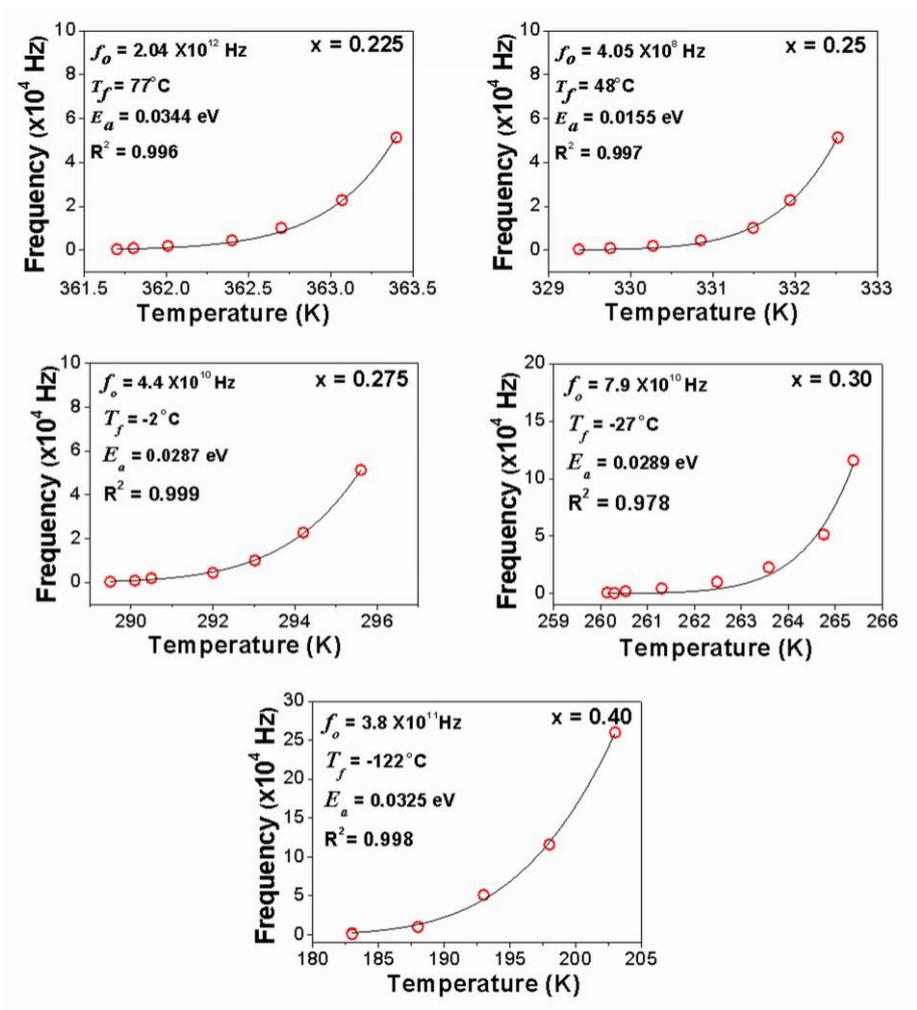

**FIGURE 6:**

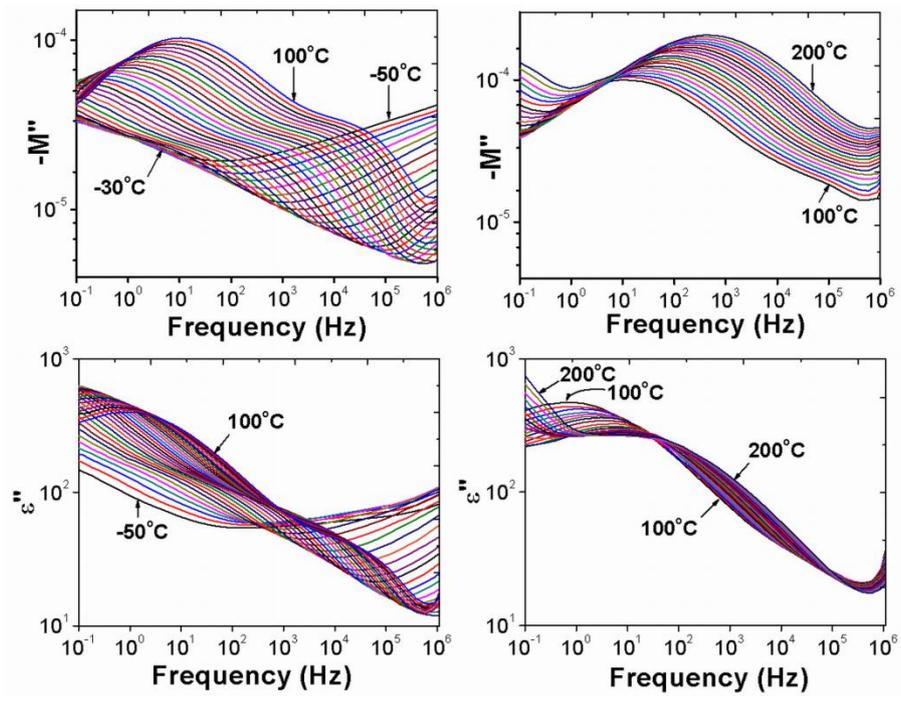